\documentclass{nature}
\usepackage[utf8]{inputenc}
\usepackage[T1]{fontenc}
\usepackage{graphicx}
\usepackage{xcolor}
\usepackage{physics}
\usepackage{bm}
\usepackage{amsfonts}
\makeatletter
\let\saved@includegraphics\includegraphics
\AtBeginDocument{\let\includegraphics\saved@includegraphics}
\renewenvironment*{figure}{\@float{figure}}{\end@float}
\makeatother

\title{Annihilation of exceptional points from different Dirac valleys  in a 2D photonic system}

\author{M.~Kr\'ol$^{1}$, I.~Septembre$^{2}$\footnote{These authors contributed equally: M.~Kr\'ol, I.~Septembre.}, P.~Oliwa$^{1}$, M.~K\k{e}dziora$^{1}$, K.~\L{}empicka-Mirek$^{1}$, M.~Muszy\'nski$^{1}$, R.~Mazur$^{3}$, P.~Morawiak$^{3}$, W.~Piecek$^{3}$, P.~Kula$^{4}$, W.~Bardyszewski$^5$, P.~G.~Lagoudakis$^{6,7}$, D.~D.~Solnyshkov$^{2,8\ast}$, G.~Malpuech$^{2\ast}$, B.~Pi\k{e}tka$^{1\ast}$, J.~Szczytko$^{1\ast}$}

\begin{document}
\maketitle

\begin{affiliations}
\item Institute of Experimental Physics, Faculty of Physics, University of Warsaw, Poland
\item Institut Pascal, PHOTON-N2, Universit\'e Clermont Auvergne, CNRS, Clermont INP, Clermont, F-63000 Clermont-Ferrand, France
\item Institute of Applied Physics, Military University of Technology, Warsaw, Poland
\item Institute of Chemistry, Military University of Technology, Warsaw, Poland
\item Institute of Theoretical Physics, Faculty of Physics, University of Warsaw, Poland
\item Skolkovo Institute of Science and Technology, Bolshoy Boulevard 30, bld. 1, Moscow, 121205, Russia
\item Department of Physics and Astronomy, University of Southampton, Southampton SO17 1BJ, UK
\item Institut Universitaire de France (IUF), F-75231 Paris, France
\end{affiliations}

\begin{abstract}
Topological physics relies on the existence of Hamiltonian's eigenstate singularities carrying a topological charge, such as quantum vortices, Dirac points, Weyl points and -- in non-Hermitian systems -- exceptional points (EPs), lines or surfaces~\cite{Bliokh2017,Shen2018,bergholtz2021exceptional}. 
They appear only in pairs connected by a Fermi arc and are related to a Hermitian singularity, such as a Dirac point. The annihilation of 2D Dirac points carrying opposite charges has been experimentally reported~\cite{Tarruell2012,Bellec2013PRL,milicevic2019type}. It remained elusive for Weyl points and second order EPs terminating different Fermi arcs.
Here, we observe the annihilation of second order EPs issued from different Dirac points forming distinct valleys. We study a liquid crystal microcavity~\cite{rechcinska2019engineering} with voltage-controlled birefringence and TE-TM photonic spin-orbit-coupling~\cite{gianfrate2020measurement}. Two neighboring modes can be described by a two-band Hermitian Hamiltonian showing two topological phases with either two same-sign or four opposite-sign Dirac points (valleys). Non-Hermiticity is provided by polarization-dependent losses~\cite{Richter2019,Liao2021}, which split Dirac points into pairs of EPs, connected by Fermi arcs. We measure their topological charges and control their displacement in reciprocal space by increasing the non-Hermiticity degree. EPs of opposite charges from different valleys meet and annihilate, connecting in a closed line the different Fermi arcs. This non-Hermitian topological transition occurs only when the Hermitian part of the Hamiltonian is topologically trivial (with four valleys), but is distinct from the Hermitian transition. 
Our results offer new perspectives of versatile manipulation of EPs, opening the new field of non-Hermitian valley-physics.
\end{abstract}

\flushbottom
\maketitle
\thispagestyle{empty}

The eigenstates of a Hamiltonian combining Hermitian and non-Hermitian parts are, in general, non-orthogonal. Exceptional points (EPs), where the eigenstates coalesce, can appear at the maxima of non-orthogonality when the relative non-Hermiticity is increased. EPs are known in optics for more than a century \cite{Voigt1902}, but only recently they have been shown to allow remarkable phenomena \cite{Miri2019}, such as specific lasing \cite{Peng2014}, unidirectional transport \cite{Doppler2016}, enhanced sensing \cite{hodaei2017enhanced,chen2017exceptional,lai2019observation,hokmabadi2019non}, or scattering control \cite{song2021plasmonic}. 
Their importance has been revealed thanks to their description in terms of a topological charge, characterizing a topological phase \cite{Bliokh2017,gong2018topological,Shen2018,bergholtz2021exceptional}, which can be measured by encircling the EP in parameter space \cite{Gao2015b,su2021direct}. Contrary to 2D Hermitian systems, topological transitions in the non-Hermitian case cannot occur through simple band touching, but only via creation or annihilation of topological charges. 
So far, the non-Hermitian topological transitions which have been reported were related to the \emph{creation} of EPs 
\cite{Zeuner2015,Ozturk2021}. EPs always appear by pair connected by a Fermi arc in the full parameter space, similar to the Weyl points in Hermitian systems, as described by the famous Nielsen-Ninomiya no-go theorem \cite{Nielsen1981}. Each pair of EPs is formed from a minimum of the Hermitian coupling (e.g. band crossing). In periodic systems, EPs are necessarily located on a closed line in parameter space and an increase of non-Hermiticity leads to the extension of the Fermi arcs which eventually form closed Fermi loops when EPs meet. More complicated situations can occur when more than two oscillators, or two bands get coupled, which leads to the emergence of higher-order singularities \cite{ding2016emergence,Tang2020}. 

In this work, we demonstrate a different type of non-Hermitian topological transition in a continuous (non-periodic) 2D photonic system. We show that if the Hermitian Hamiltonian is topologically trivial, the EPs issued from different Hermitian singularities and limiting different Fermi arcs can meet and \emph{annihilate} upon the \emph{increase} of non-Hermiticity.

\begin{figure}[tbp]
\centering
\includegraphics[width=0.9\linewidth]{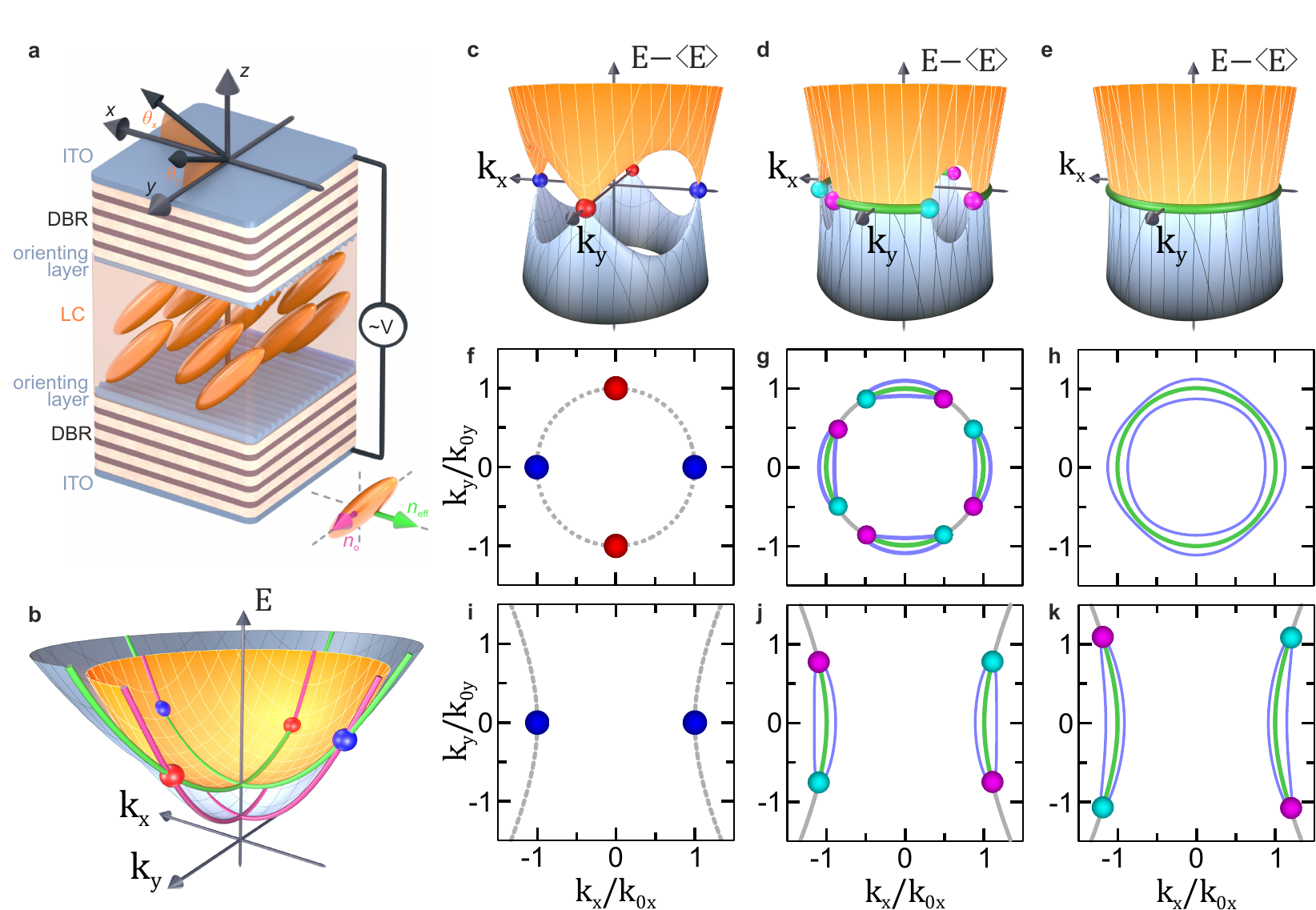}

\caption{\textbf{Scheme of the experiment and the possible behaviors of the exceptional points.} \small
\textbf{a} Distributed Bragg Reflector (DBR) based microcavity filled by liquid crystal molecules whose orientation is controlled by an external voltage. \textbf{b} 2D Energy dispersion of the mode $N+2$ polarized H at $k=0$ and of the mode $N$ polarized V at $k=0$, $k_i=(\omega \sin\theta_i)/c$. \textbf{c}-\textbf{e} Difference of the real part of the energies when $\beta'>\beta$ and in the Hermitian case \textbf{c}, and non-Hermitian cases for $2\Delta=3$~meV \textbf{d} and $2\Delta=1.2$~meV \textbf{e}. Other used parameters are given in the main text. \textbf{f}-\textbf{h} Same as \textbf{c}-\textbf{e} but view from the top. The points sets the positions in $k$-space of the four Dirac points \textbf{f} and eight EPs \textbf{g}, whereas their colors marks the sign of their topological charges. The dashed line in \textbf{f} shows the ellipse given by Eq.~\eqref{ellipse}, setting the allowed positions of EPs. EPs coordinates are determined by the crossing of the ellipse and the blue lines given by $\bm{\Omega}_r^2-\bm{\Omega}_i^2=0$. This crossing breaks the ellipse in Fermi arcs shown in green in \textbf{g},\textbf{h} and imaginary Fermi arcs shown in gray in  \textbf{g}. \textbf{i}-\textbf{k} Same as \textbf{f}-\textbf{g} but when  $\beta'<\beta$ and the Hermitian limits contains only two same sign Dirac points. EPs from different Fermi arcs \textbf{j},\textbf{k} cannot annihilate, belonging to two distinct separated trajectories.}
\label{fig1}
\end{figure}

The system we study is a microcavity filled with a liquid crystal\cite{rechcinska2019engineering,krol2021observation} (Fig.~\ref{fig1}a). This Fabry-Perot resonator
hosts a series of photonic modes with quantized wave vectors  perpendicular to the mirror plane and energies $E_N$ ($N$ is the mode number). Each mode forms a polarization doublet showing an in-plane parabolic dispersion with a 2D effective mass which scales as $N$. The polarization degeneracy in a doublet is lifted at all wave-vectors except at $k=0$ (touching parabolas) by the splitting between TE and TM eigenmodes  (Transverse-Electric and Transverse-Magnetic). This splitting acts as photonic spin-orbit coupling characterized by a winding number 2 \cite{Kavokin2005,gianfrate2020measurement}. The liquid crystal molecules orientation is set by an external voltage, which controls a tunable linear birefringence $\alpha$ \cite{rechcinska2019engineering,krol2021observation}. A small birefringence  $\alpha<(E_{N+1}-E_N)$ lifts the $k=0$ degeneracy, leading to the formation of two tilted Dirac cones both carrying the same topological Berry charge\cite{gianfrate2020measurement} $+1/2$. When $\alpha$ becomes comparable with $(E_{N+1}-E_N)$, modes of different parities become energetically close and get coupled by a Rashba-Dresselhaus spin-orbit coupling with
equal strength \cite{rechcinska2019engineering}, also called emergent optical activity \cite{Ren2021}. Here we consider $\alpha\approx (E_{N+2}-E_N)$ \cite{krol2021observation}, so that this optical activity is negligible. The corresponding eigenmodes look like two 2D parabola (Fig.~\ref{fig1}b). Neglecting the decaying character of modes, these two bands can be described by the following effective $2\times 2$ Hermitian Hamiltonian written on the circular polarization basis:
\begin{equation}
H_{\bf{k}}^{real} = \left( {\begin{array}{*{20}{c}}
{\frac{{E_H^{N+2} + E_V^{N }}}{2} + \frac{{{\hbar ^2}k_x^2}}{{2{m_x}}} + \frac{{{\hbar ^2}k_y^2}}{{2{m_y}}}}&{\Delta  - \beta '{k^2} - \beta {{({k_x} - i{k_y})}^2}}\\
{\Delta  - \beta '{k^2} - \beta {{({k_x} + i{k_y})}^2}}&{\frac{{E_H^{N+2} + E_V^{N }}}{2} + \frac{{{\hbar ^2}k_x^2}}{{2{m_x}}} + \frac{{{\hbar ^2}k_y^2}}{{2{m_y}}}}
\end{array}} \right)
\end{equation}
where $E_H^{N+2}$ and $m_H\sim N+2$ are the energy and mass, of the $N+2$th H-polarized mode and $E_V^{N}$ and $m_V\sim N$, are the energy and mass of the V-polarized mode number $N$. $k_x$, $k_y$ are the 2D wave vector components.
The spin-independent masses $m_x$ and $m_y$ are determined by the birefringence and the angle of the optical axis (see Supplementary). $\beta$ is the magnitude of the TE-TM spin orbit coupling. $\beta'=\hbar^2(m_H-m_V)/4m_Hm_V$ and $\Delta=(E_H^{N+2}-E_V^{N})/2$.
This Hermitian Hamiltonian can be written as a linear combination of identity and Pauli matrices, which defines a real effective magnetic field $\bm{\Omega_r}$ acting on the polarization pseudo-spin ($H=\bm{\Omega_r}\cdot\bm{S}$). The two non-zero components of the field are $\Omega_r^x=\Delta-\beta'k^2-\beta(k_x^2-k_y^2)$ and  $\Omega_r^y=-2 \beta k_x k_y$.
 
This effective Hamiltonian possesses two distinct topological phases described and observed  in the Supplementary Materials. If $\beta>\beta'$, the bands show two tilted Dirac cones  carrying the same topological charge (Fig.~\ref{fig1}i), and the bands are  topologically non-trivial, characterized by a non-zero winding of the pseudospin giving rise to a non-zero Chern number, if a gap is opened by breaking the time-reversal symmetry. If  $\beta<\beta'$, which is the case in our experiment, the bands possess four tilted Dirac cones, as shown in Fig.~\ref{fig1}c,f. Their coordinates are given by $(\pm{k_{0x}},0)$ and $(0,\pm{k_{0y}})$,
where $k_{0,x,y}={(\Delta/(\beta'\pm\beta))^{1/2}}$. The winding number of the pseudospin is $+1$ for the two Dirac points located on the $k_y$-axis, as previously, which corresponds to Berry curvature monopoles of charge $+1/2$. The Dirac points located on the $k_x$-axis carry a pseudospin winding number $-1$ and a Berry curvature charge $-1/2$ (Fig.~\ref{fig1}f). The corresponding bands are therefore globally topologically trivial, with zero overall pseudospin winding and a vanishing integral of the Berry curvature (Chern number).

The next step is to consider the losses (line broadening), inherently present in photonic systems. Importantly, these losses are significantly different for the H and V modes with $\Gamma_H=2.04$~meV, $\Gamma_V=1.8$~meV (see Supplementary Materials).  This requires adding an imaginary (non-Hermitian) part to the total Hamiltonian of the system as already illustrated in \cite{Richter2019}:
\begin{equation}
H_{\bf{k}} = H_{\bf{k}}^{real}+H^{imag}
\label{HNH}
\end{equation}
where 
$H^{imag}=i(\Gamma_0\mathbb{I}_2+\delta\Gamma\sigma_x)$
with $\Gamma_0=(\Gamma_H+\Gamma_V)/2$ is the mean decay rate and $\delta\Gamma=(\Gamma_H-\Gamma_V)/2$ defines a constant imaginary effective field along $x$: $\bm{\Omega}_i=(\delta\Gamma,0,0)^T$. As shown in Fig.~\ref{fig1}d,g, this non-Hermitian part transforms each Dirac point into a pair of EPs \cite{Voigt1902,Richter2019,Liao2021} connected by a line, called a Fermi arc \cite{bergholtz2021exceptional}, where the real parts of the eigenvalues are degenerate. The squared absolute value of the \emph{complex} splitting between the eigenmodes reads $4(|\bm{\Omega}_r^2-\bm{\Omega}_i^2|^2+4|\bm{\Omega}_r\bm{\Omega}_i|^2)$.
The existence of an exceptional point (zero splitting) therefore requires 
 $\bm{\Omega}_r\bm{\Omega}_i=0$ and $\bm{\Omega}_r^2-\bm{\Omega}_i^2=0$.
In our case, the first condition reads:
\begin{equation}
\frac{k_x^2}{k_{0x}^2}+\frac{k_y^2}{{k_{0y}^2}}=1
\label{ellipse}
\end{equation}
which determines an ellipse of possible location for EPs (Fig.~\ref{fig1}g, cyan and magenta points). The second condition $\bm{\Omega}_r^2-\bm{\Omega}_i^2=0$ is verified along the blue curves in Fig.~\ref{fig1}g,h. The crossing of both lines sets the coordinates of the 8 EPs which read:
\begin{eqnarray}
k_y^e & = & \pm \frac{k_{0y}}{\sqrt{2}}\sqrt{{1\pm\sqrt{1-{\delta\Gamma}^2/{\beta^2 k_{0x}^2k_{0y}^2}}}},\\
k_x^e & = & \pm \frac{k_{0x}}{\sqrt{2}}\sqrt{{1\pm\sqrt{1-{\delta\Gamma}^2/{\beta^2 k_{0x}^2k_{0y}^2}}}}.
\label{EPcoords}
\end{eqnarray}
The Fermi arc, shown in blue-green, is a part of the ellipse where $\bm{\Omega}_r^2-\bm{\Omega}_i^2<0$, whereas the other part, given by $\bm{\Omega}_r^2-\bm{\Omega}_i^2>0$ and shown in gray is the imaginary Fermi arc\cite{bergholtz2021exceptional} with degenerate imaginary parts of the modes.

\begin{figure}[tbp]
\centering
\includegraphics[width=0.99\linewidth]{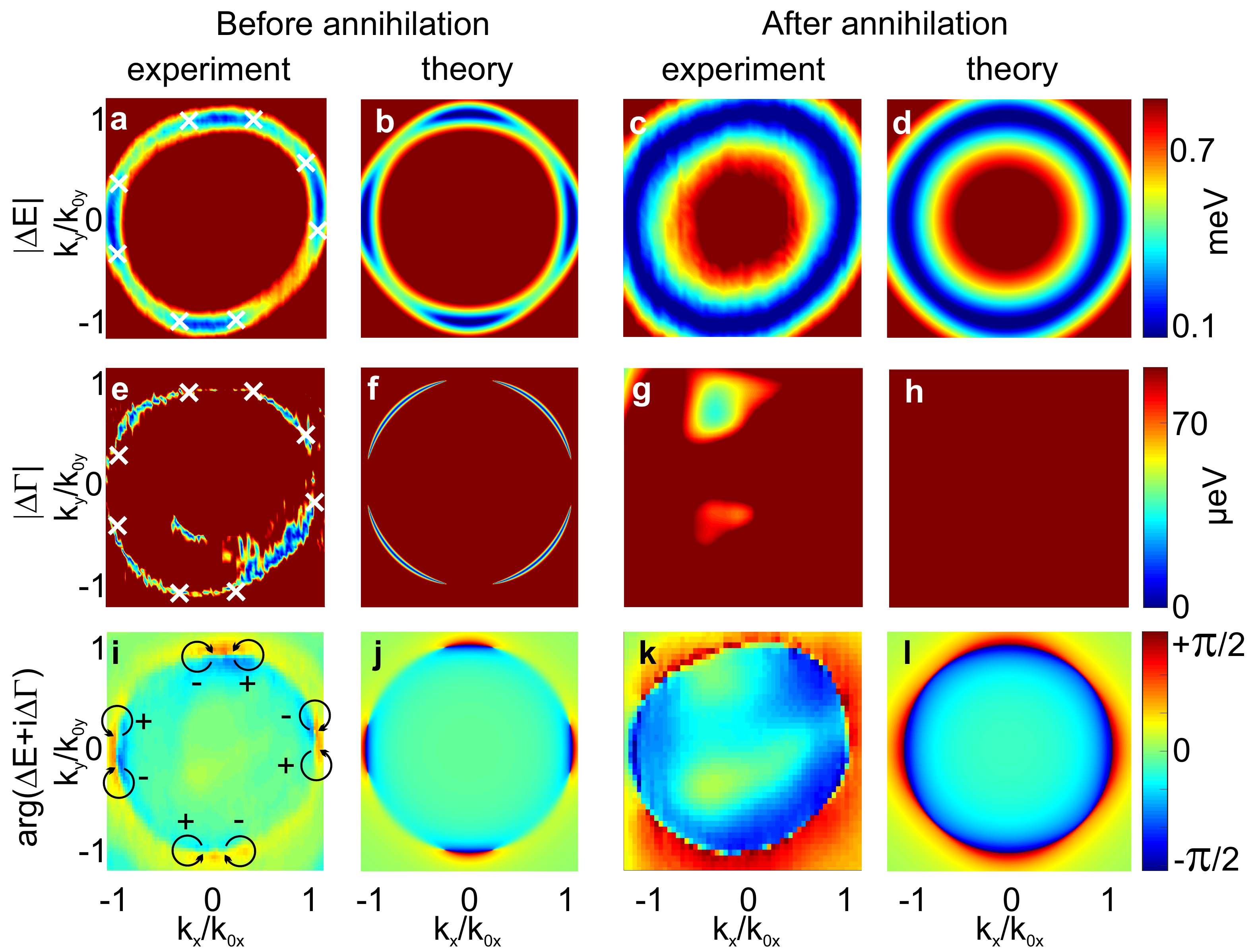}
\caption{\textbf{Observation of EPs, of their topological charges and of their annihilation} \small
Theoretical figures are obtained by diagonalizing the Hamiltonian (2) with parameters given in Methods. The procedure for the extraction experimental energies is detailed in  Methods. \textbf{a}-\textbf{d} Difference of the real part of the eigenenergies for $2\Delta = 3$~meV (\textbf{a},\textbf{b}) and $1.2$~meV (\textbf{b},\textbf{d}). The color bar is saturated above $0.9$~meV.  The white crosses in \textbf{a} shows the EP coordinate limiting the blue area corresponding to Fermi arcs (degenerate real part of energy). \textbf{e}-\textbf{h} same as \textbf{a}-\textbf{d}, but for imaginary part.  The color bar is saturated above $90~\mu$eV. Imaginary Fermi arcs, where imaginary parts of energies are degenerate, appear in blue.
\textbf{i,k} Experimental  and \textbf{j,l} theoretical  phase (argument) of the difference of complex energies. EPs are associated to a vortex phase with a phase shift $\pm\pi$ whose winding is shown by the arrows in \textbf{i}.} 
\label{fig2}
\end{figure}

The topological charge of an EP can be defined as the winding number of the complex energy of eigenmodes around the EP \cite{Bliokh2017,gong2018topological,Shen2018}:
\begin{equation}
    w=\frac{1}{2\pi}\oint d\bm{k}\cdot \bm{\nabla}_{\bm{k}} \arg E_n(\bm{k}).
\end{equation}
The winding numbers of the eight EPs alternate in sign along the ellipse. Increasing $\delta\Gamma$ or decreasing $\Delta$ increases the degree of non-Hermiticity and moves the exceptional points away from the spawning points along the ellipse, until they meet each other and annihilate, as shown in Fig.~\ref{fig1}h. Fermi arcs connect to form a closed line of trivial degeneracy. We note that it is not a ring of exceptional points reported in \cite{zhen2015spawning}, because the imaginary parts of the energies are not degenerate along this whole line.  

As said above, a topological transition occurs for the Hermitian part of the Hamiltonian when $\beta'=\beta$. The system switches between four Dirac points (globally trivial) and two Dirac points (non-trivial, Fig.~\ref{fig1}i). This topological transition is demonstrated experimentally in Supplemental Materials. In this case, $k_{0y}$ becomes imaginary. Equation~\eqref{ellipse} determining the location of EPs remains valid, but describes two hyperbolas, as shown in Fig.~\ref{fig1}j,k. EPs issued from distinct Dirac points are moving towards infinity on separated open curves and cannot meet anymore. In such a case, the non-Hermitian topological transition associated with the annihilation of the exceptional points cannot occur. This situation is impossible in periodic systems, where the Fermi lines are necessarily always closed because the periodic energy dispersion is necessarily finite. In continuous systems, such as optical cavities, this situation is typically realized for weak birefringence \cite{gianfrate2020measurement}, $E_H^N\approx E_V^N$. 

Going back to the case $\beta<\beta'$, the non-Hermitian topological transition associated with EP annihilation occurs in Eqs.~\eqref{EPcoords} when $\delta\Gamma=\beta k_{0x} k_{0y}$, which gives
\begin{eqnarray}
\frac{\delta\Gamma}{\Delta}={\frac{\beta}{\sqrt{\beta'^2-\beta^2}}}.
\end{eqnarray}
One can see that the degree of non-Hermiticity can be changed either by increasing  $\delta\Gamma$ or decreasing $\Delta$. The latter option is used in our experiment: $\Delta$ is controlled by the voltage affecting the liquid crystal molecules orientation. 
The experimental study of the liquid crystal cavity is performed by polarization-resolved reflectance, from which we extract the real and imaginary parts of the energies of the eigenmodes and also their polarization (see Methods). The difference between the real (imaginary) part of the energies versus $k_x,k_y$ are shown in Fig.~\ref{fig2}a,e and Fig.~\ref{fig2}c,g for two values of detuning $2\Delta =3$~meV and $1.2$~meV for the cases where the EPs are present and annihilated, respectively. In Fig.~\ref{fig2}a,e, one can observe the Fermi arcs, where the real parts of energy are degenerate (within the measurement accuracy) and the imaginary parts split. The EPs are shown by white crosses. On the other hand, imaginary Fermi arcs are the lines where the imaginary parts of energies are degenerate and the real parts split. After the EPs annihilation, only a real Fermi arc remains (Fig.~\ref{fig2}c,g). We note that in Ref.\cite{krol2021observation}, the very same structure was experimentally studied in the same regime of crossing between the $N+2$ and $N$ modes. The key difference is that the experiment was performed at higher detuning $\Delta$, so that Fermi arcs were expected to be very small, below the grid size in k-space, and thus the EPs could not be resolved. Figure~\ref{fig2}b,d,f,h shows the corresponding theoretical results, obtained using the effective non-Hermitian Hamiltonian~\eqref{HNH}. 

The topological charge measurement is presented in Fig.~\ref{fig2}i-l.
Panels~i,k show a map of the phase of the complex energy of the lower mode for $2\Delta=3$~meV and $2\Delta=1.2$~meV, as previously. In both cases, the real Fermi arcs appear as a sharp phase shift. Very clear phase vortices are visible at the EPs positions in Fig.~\ref{fig2}i and are absent in Fig.~\ref{fig2}k. These features are in excellent agreement with the simulations based on the effective non-Hermitian Hamiltonian~\eqref{HNH}, shown in Fig.~\ref{fig2}j,l. The topological charge of each EP $w=\pm1/2$ is determined by the direction of the phase vortex winding. These charges are opposite for the EPs originating from different Dirac points, which ultimately allows their annihilation.

\begin{figure}[tbp]
\centering
\includegraphics[width=0.75\linewidth]{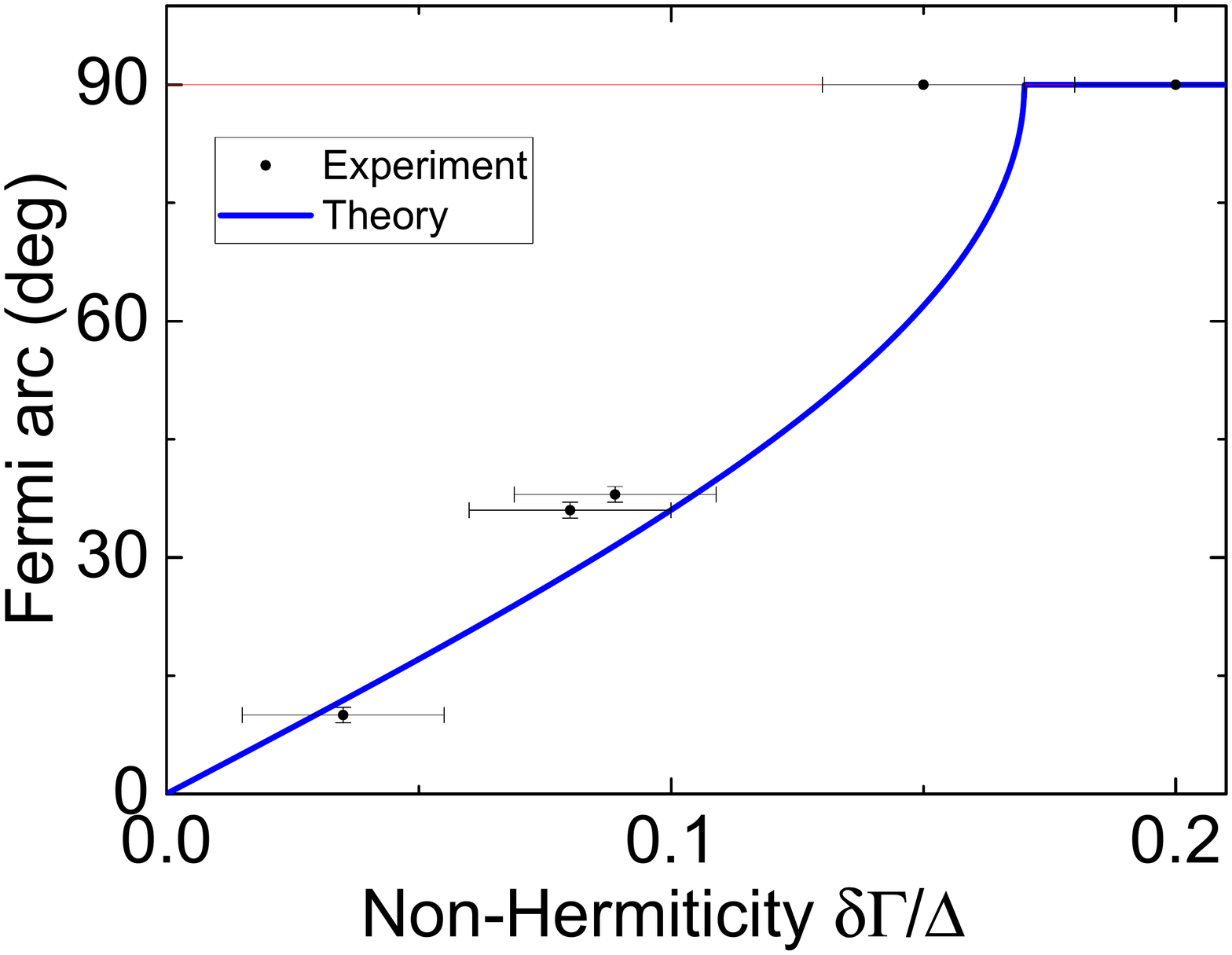}
\caption{\textbf{Moving and annihilating exceptional points.} The length of the Fermi arc as a function of non-Hermiticity. Black points -- experiment, blue line -- theory. The maximal length of each Fermi arc (90$^\circ$) is marked with a thin red line. Error bars indicate the measurement uncertainty. The experimental points are obtained by varying the non-Hermiticity degree $\delta\Gamma/\Delta$ by controlling the detuning $\Delta$ via applied voltage $V$.}
\label{fig3}
\end{figure}

Figure 3 demonstrates the control of the Fermi arc angular size related to 
the position of EPs through the experimental tuning of the non-Hermiticity parameter, relate to the detuning. The topological transition associated to the EPs annihilation is clearly visible taking place around $\frac{\delta\Gamma}{\Delta}$=0.17.
 
Non-Hermitian transitions in two-band systems through EP merging are typically related to a single Hermitian singularity. \cite{Bliokh2017,Shen2018,bergholtz2021exceptional}. Here, we consider the merging of EPs originating from different Dirac points upon \emph{increasing} the non-Hermiticity, which to our knowledge was not reported before, and can be viewed as a first example of non-Hermitian multi-valley physics. Interestingly, the transition we observe could occur only because the Hermitian part of the Hamiltonian is globally topologically trivial. The sum of the topological charges over a band is zero which translates into the presence of a closed line for Fermi arcs. On the contrary, when Hermitian bands are topologically non-trivial, such non-Hermitian topological transition cannot occur. The Hermitian and non-Hermitian transitions allowed by the Hamiltonian representing our system are clearly distinct. On the other side, the possibility to observe the non-Hermitian transition is entirely related to the topology of the Hermitian part. From an applied perspective, our study demonstrate the tuning of the EP coordinates in k-space by simple modification of an external voltage, in a micro-device, at optical frequencies. This ultimately allows to control the angle of emission for the associated chiral modes. Another interest of the planar cavity platform we are using is in that it allows implementing interacting photons modes (exciton-polaritons), up to the blockade regime \cite{Zasedatelev2021}. This possibility combined with our present finding could allow to address topological non-Hermitian physics for interacting particles, up to strongly correlated regimes. \cite{Zasedatelev2021}.

\begin{methods}

\textit{Sample}
The cavity consists of two distributed Bragg reflectors made of 6 SiO$_2$/TiO$_2$ pairs with maximum reflectance at 550\,nm grown on glass plates with ITO electrodes. Space between the DBRs is filled with highly birefringent liquid crystal with $\Delta n = 0.41$. 
To obtain homogeneous orientation of LC both DBRs are finished with structured polymer orienting layer. The total thickness of the LC layer is approx. 1.8\,$\mu$m.

\textit{Experimental setup}
Experimental results were obtained in transmission configuration. Broadband light from a LED diode was focused on the sample with a microscope objective with 50$\times$ magnification and numerical aperture $\text{NA} = 0.6$. Transmitted light was collected by another objective with 20$\times$ magnification and $\text{NA} = 0.4$. Fourier plane of the collecting objective was imaged on the entrance slit of a monochromator equipped with a CCD camera. Both wave vector directions were measured by scanning of the image across the slit by the automated movement of the imaging lens. Data was collected independently for 6 incident light polarizations; linear: horizontal, vertical, diagonal, antidiagonal and circular $\sigma^+$, $\sigma^-$ by adjusting angles of half wave plate and quarter wave plate after fixed linear polarizer.

LC layer anisotropy in $x$--$z$ plane was controlled by external square waveform applied to ITO electrodes with 1\,kHz frequency and amplitude of 1.77\,V (Fig.~2a,e,i) and 1.72\,V (Fig.~2c,g,k).

\textit{Linewidth extraction}
We extract the real and imaginary parts of the energies (that is, the positions and the linewidths) of the modes from the polarization-resolved spectra by fitting them with the Voigt function, in order to account both for the homogeneous broadening due to the mode lifetime and for the inhomogeneous broadening due to disorder. Only the homogeneous part of the broadening (Lorentzian linewidth) can give rise to non-Hermiticity and is accounted for by the Hamiltonian \eqref{HNH}. Parallel computing is used to speed up the extraction for the whole reciprocal space with high resolution. An example of the energy spectrum in two polarizations, together with its fit, is provided in the Supplemental Materials.

\textit{Parameters of the Hamiltonian}
The theoretical panels of Fig.~2 were calculated using the following parameters, obtained from fitting the dispersion and from the linewidth extraction discussed above: $m_x=(1.34\pm0.07)\times10^-5m_0$, $m_y=(1.08\pm0.06)\times10^-5 m_0$, $\beta=0.080\pm0.03$~meV$\mu$m$^2$, $\beta'=0.47\pm0.03$~meV$\mu$m$^2$. Other parameters were already given in the text, but we provide them here for convenience: $2\Delta=2.7\pm0.1$ and $1.2\pm0.1$~meV, $2\delta\Gamma=0.24\pm0.05$~meV.

\end{methods}

\bibliographystyle{naturemag}
\bibliography{biblio}

\begin{addendum}
\item We acknowledge useful discussions with C. Leblanc.
This work was supported by the National Science Centre grants 2019/35/B/ST3/04147, 2019/33/B/ST5/02658 and 2017/27/B/ST3/00271, and the Ministry of National Defense Republic of Poland Program -- Research Grant MUT Project 13-995 and MUT University grant (UGB) for the Laboratory of Crystals Physics and Technology for year 2021, and the European Union's Horizon 2020 program, through a FET Open research and innovation action under the grant agreements No. 899141 (PoLLoC) and No. 964770 (TopoLight). We also acknowledge the support  of the ANR Labex Ganex (ANR-11-LABX-0014), and of the ANR program "Investissements d'Avenir" through the IDEX-ISITE initiative 16-IDEX-0001 (CAP 20-25). P.G.L. acknowledge the support of the UK's Engineering and Physical Sciences Research Council (grant EP/M025330/1 on Hybrid Polaritonics), the support of the RFBR project No. 20-52-12026 (jointly with DFG) and No. 20-02-00919. 
\item[Competing interests] The authors declare no competing interests. 
\item[Correspondence] Correspondence related to the experiments should be addressed to Jacek.Szczytko@fuw.edu.pl (JS); Barbara.Pietka@fuw.edu.pl (BP); Correspondence related to theory should be addressed to dmitry.solnyshkov@uca.fr (DS); guillaume.malpuech@uca.fr (GM). 
\end{addendum}

\section*{Data availability statement}
The datasets generated during and/or analysed during the current study are available in the Open Science Framework (OSF) repository,

\verb+https://osf.io/jnx8k/?view_only=16426f9a35404264badaaa93162060a7+



\renewcommand{\thefigure}{S\arabic{figure}}
\setcounter{figure}{0}
\renewcommand{\theequation}{S\arabic{equation}}
\setcounter{equation}{0}

\section*{Supplementary Information}
In this Supplementary Information, we provide more details on the behavior of the Hermitian part of the Hamiltonian of our system. We demonstrate how the Hamiltonian parameters are extracted from the experiment, including the non-Hermiticity (linear dichroism). Finally, we demonstrate that the winding of the real effective field constrains the possibility of a non-Hermitian topological transition with annihilation of the exceptional points: The annihilation is only possible if the total winding of the real effective field is zero.

\subsection{Topology of the Hermitian Hamiltonian: Theory}

In this section, we illustrate the topological transitions concerning the Hermitian part of the two-band Hamiltonian (1) of the main text describing the intersection of cavity modes with different numbers. This Hamiltonian is shown below for convenience, together with the definitions:
\begin{equation}
H_{\bf{k}}^{real} = \left( {\begin{array}{*{20}{c}}
{\frac{{E_H^{N+2} + E_V^{N}}}{2} + \frac{{{\hbar ^2}k_x^2}}{{2{m_x}}} + \frac{{{\hbar ^2}k_y^2}}{{2{m_y}}}}&{\Delta  - \beta '{k^2} - \beta {{({k_x} - i{k_y})}^2}}\\
{\Delta  - \beta '{k^2} - \beta {{({k_x} + i{k_y})}^2}}&{\frac{{E_H^{N+2} + E_V^{N}}}{2} + \frac{{{\hbar ^2}k_x^2}}{{2{m_x}}} + \frac{{{\hbar ^2}k_y^2}}{{2{m_y}}}}
\end{array}} \right),
\end{equation}
where $E_H^{N+2}$ and $m_H\sim N+2$ are the energy and mass, of the $N+2$th H-polarised mode and $E_V^{N}$ and $m_V\sim N$, are the energy and mass of the V-polarised mode number $N$. $k_x$, $k_y$ are the 2D wave vector components.
The spin-independent masses $m_x$ and $m_y$ are determined by the birefringence ($n_e$, $n_o$) and the angle of the optical axis $\theta$, according to the following expressions \cite{rechcinska2019engineering}: 
\begin{eqnarray}
    \frac{1}{m_x}&=&\frac{1}{2m'_0}\left(\frac{n_o^2\sin^2\theta+n_e^2\cos^2\theta+n_e^2}{n_e^2 n_o^2}\right),\\
    \frac{1}{m_y}&=&\frac{1}{2m'_0}\left(\frac{n_e^2+n_o^2}{n_e^2 n_o^2}\right),
\end{eqnarray}
where $m'_0$ is the mass at the central frequency (between the two modes in question).
$\beta$ is the magnitude of the TE-TM spin orbit coupling. $\beta'=\frac{\hbar^2(m_V-m_H)}{2m_Hm_V}$ and $\Delta=(E_H^{N+2}-E_V^{N})/2$.
This Hermitian Hamiltonian can be written as a linear combination of identity and Pauli matrices which defines a real effective magnetic field $\Omega_r$ acting on the polarization pseudo-spin. The two non-zero components of the field are $\Omega_r^x=\Delta-\beta'k^2-\beta(k_x^2-k_y^2)$ and  $\Omega_r^y=-2 \beta k_x k_y$.

The pseudospin of a given eigenstate is either aligned or anti-aligned with the effective field (so far as the Hamiltonian remains Hermitian). Experimentally, this pseudospin corresponds to the Stokes vector of light, calculated using the following expressions:
\begin{eqnarray}
    S^1&=&\frac{I_V-I_H}{I_H+I_V},\\
    S^2&=&\frac{I_D-I_A}{I_D+I_A},\nonumber\\
    S^3&=&\frac{I_R-I_L}{I_R+I_L},\nonumber
\end{eqnarray}
where the $I$s are the intensities in each of the polarizations, which can be measured experimentally or calculated theoretically as the absolute value squared of the eigenstate wavefunction in the corresponding basis. Each pseudospin component is an average value of the corresponding spin operator represented by a Pauli matrix. 

The case $\beta'=0$ has already been considered in \cite{Tercas2014,gianfrate2020measurement}. The two parabolas are split at $k=0$, with different possible signs of this splitting (band detuning).

\begin{figure}
    \centering
    \includegraphics[width=0.95\linewidth]{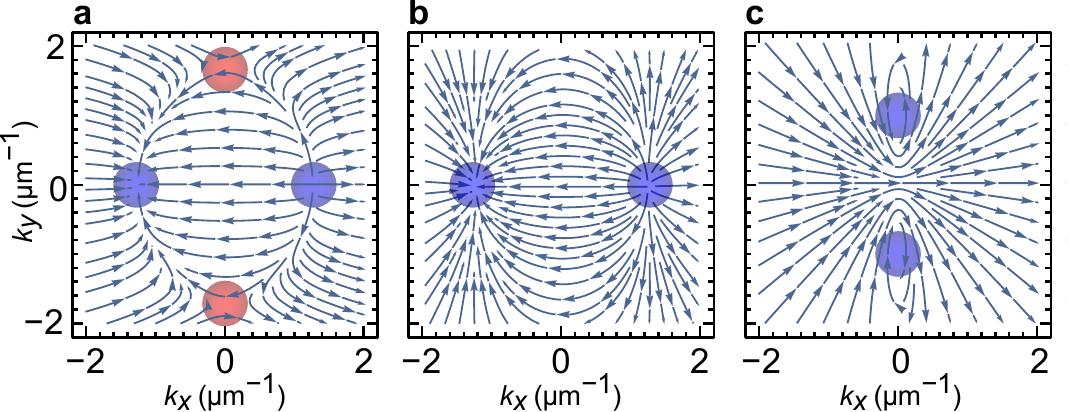}
    \caption{\textbf{Pseudospin textures for the Hermitian Hamiltonian}: \textbf{a}  trivial case $\beta<\beta'$, \textbf{b,c}  topologically non-trivial case $\beta>\beta'$ with positive and negative detunings.  The red marks shows the Dirac points of winding number +1 and the blue marks the Dirac points of winding number -1} 
    \label{texture}
\end{figure}

For positive detuning, the TE-TM splitting compensates this splitting along $k_x$ leading to a linear crossing. 
On the contrary, the TE-TM splitting is changing sign along $k_y$ and adds to the $k=0$ splitting. As a result, along $k_y$ the two parabola get away the one with respect to the other and do not cross. The resulting spectrum shows two tilted Dirac cones at $\pm k_{0x}$. The winding number of the effective field around these two points is the same (+1), as shown in Fig.~{\ref{texture}b} showing the distribution of the pseudospin (and effective field) texture in the reciprocal space. This winding is inherited from the winding number 2 of the TE and TM modes. The Berry topological charge is $+1/2$. This topological phase remains while $\beta'$ increases up to $\beta'=\beta$ where two new crossing points between the parabola appear at $k_{0y}$ which sets at infinity when the transition occurs. These crossing points correspond to tilted Dirac cones. The winding number of the in-plane pseudo-spin around these two points is (-1) and the Berry topological charge (-1/2), as shown on Figure~\ref{texture}-b. The sum of the topological charges of the 4 Dirac points is zero and the bands are overall topologically trivial which allows the non-Hermitian topological transition we explore in the main text.

For negative detuning (Fig.~\ref{texture}c), the winding of the two Dirac points does not change, but their positions in reciprocal space are rotated by $\pi/2$ and the particular field distribution changes from monopolar to Rashba.

\subsection{Topology of the Hermitian Hamiltonian: Experiment}

\begin{figure}[tbp]
    \centering
    \includegraphics[width=0.8\linewidth]{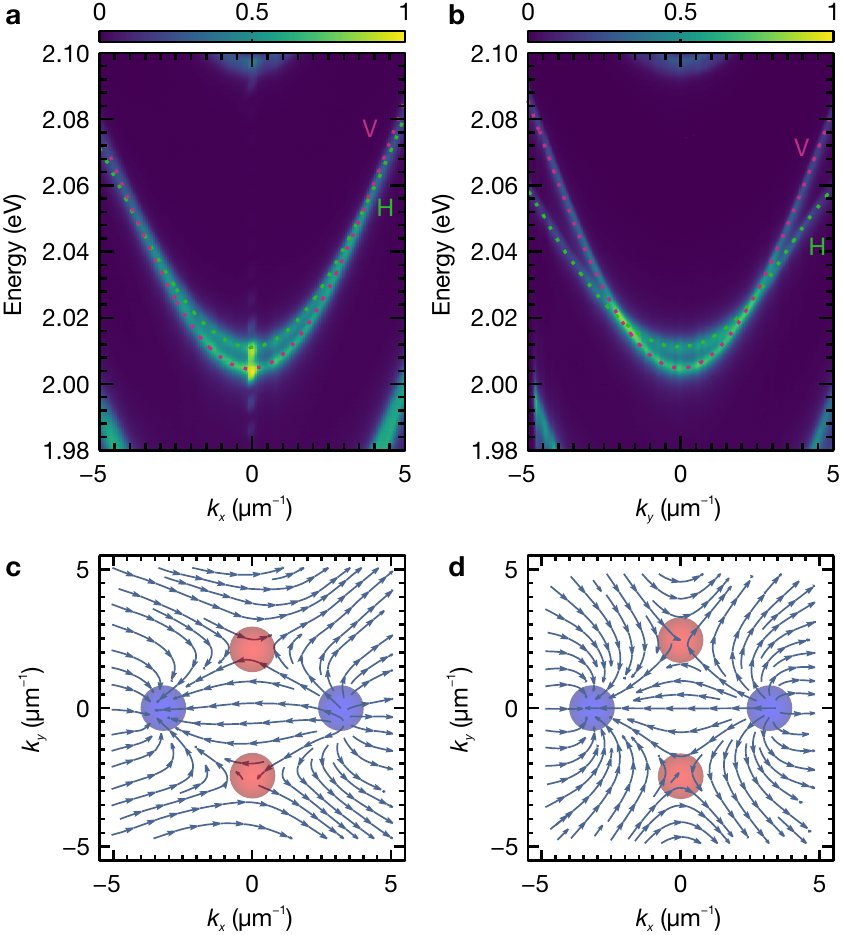}
    \caption{\textbf{4 diabolical points case}. Experimental dispersion relation along \textbf{a} $k_x$ and \textbf{b} $k_y$ directions observed at 1.39\,V applied to the microcavity, corresponding to the case with total winding 0. Energy of the bands visible in horizontal (vertical) polarization is marked by green (purple) dashed line. \textbf{c} Experimental and \textbf{d} simulated by Berreman method pseudospin texture in $S_1$--$S_2$ plane of lower energy band.}
    \label{exp_N2N}
\end{figure}

\begin{figure}[tbp]
    \centering
    \includegraphics[width=0.8\linewidth]{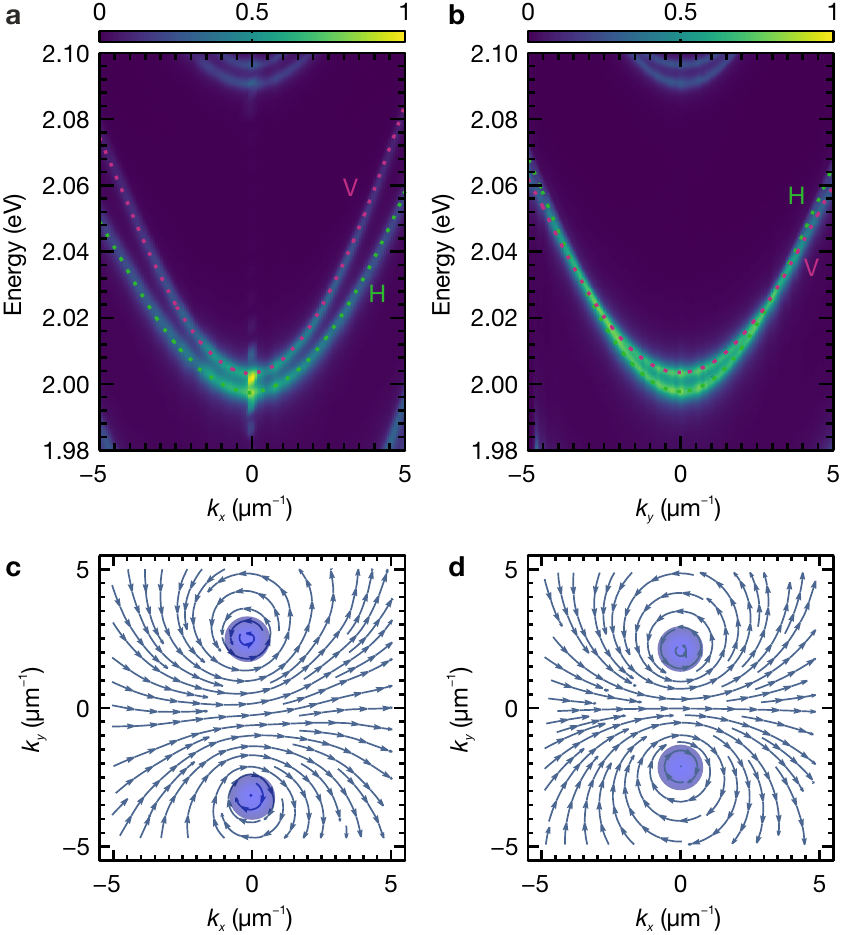}
    \caption{\textbf{2 diabolical points case}. Experimental dispersion relation along \textbf{a} $k_x$ and \textbf{b} $k_y$ directions, observed at 11\,V applied to the microcavity, corresponding to the case with winding 2. Energy of the bands visible in horizontal (vertical) polarization is marked by green (purple) dashed line. \textbf{c} Experimental and \textbf{d} simulated by Berreman method pseudospin texture in $S_1$--$S_2$ plane of lower energy band.}
    \label{exp_NN}
\end{figure}

Cavities filled with birefringent nematic liquid crystal allows for realization of both topologically-distinct cases mentioned above: with non-trivial total winding 2 and  trivial total winding 0 (discussed in the main text), depending on voltage applied to the cell. For large enough energy detuning between the two modes of interest, the relative non-Hermiticity is weak enough so that the measured pseudo-spin is simply aligned with the real effective magnetic field (Hermitian limit). In such a case, the winding of the real field in reciprocal space can be determined experimentally through polarization resolved tomography.

Experiments described in this section were preformed on a sample analogous to the one described in the main text, but with wider LC layer of around 3.9\,$\mu$m.

Fig.\,\ref{exp_N2N} presents realization of the case with winding 0, discussed in detail in the main text. Fig.\,\ref{exp_N2N}a,b presents experimental dispersion relation along $k_x$ and $k_y$ directions observed with 1.39\,V applied to ITO electrodes. At this voltage, the numbers of the modes differ by 2, and both modes cross each other along both wave vector directions.  Experimentally determined pseudospin texture in $S_1$--$S_2$ plane of the lower energy band is plotted in Fig.\,\ref{exp_N2N}c. It  matches the Berreman matrix model presented in Fig.\,\ref{exp_N2N}d, showing 0 total winding.

With amplitude of external voltage of 11\,V modes with the same mode numbers are close to degenerate, as shown on experimental dispersion relations in Fig.\,\ref{exp_NN}. In this case modes crosses each other only along $k_y$ direction. Resulting  pseudospin texture in $S_1$--$S_2$ plane shows two 2D monopoles with total winding number of 2 in both experiment Fig.\,\ref{exp_NN}c and Berreman model simulations Fig.\,\ref{exp_NN}d.

The Dirac points appear along the Y axis because of the negative value of the detuning in the experiment, but their winding is that of the TE-TM field, as expected.

\subsection{Hamiltonian parameters}

\begin{figure}[tbp]
\includegraphics[width=0.95\linewidth]{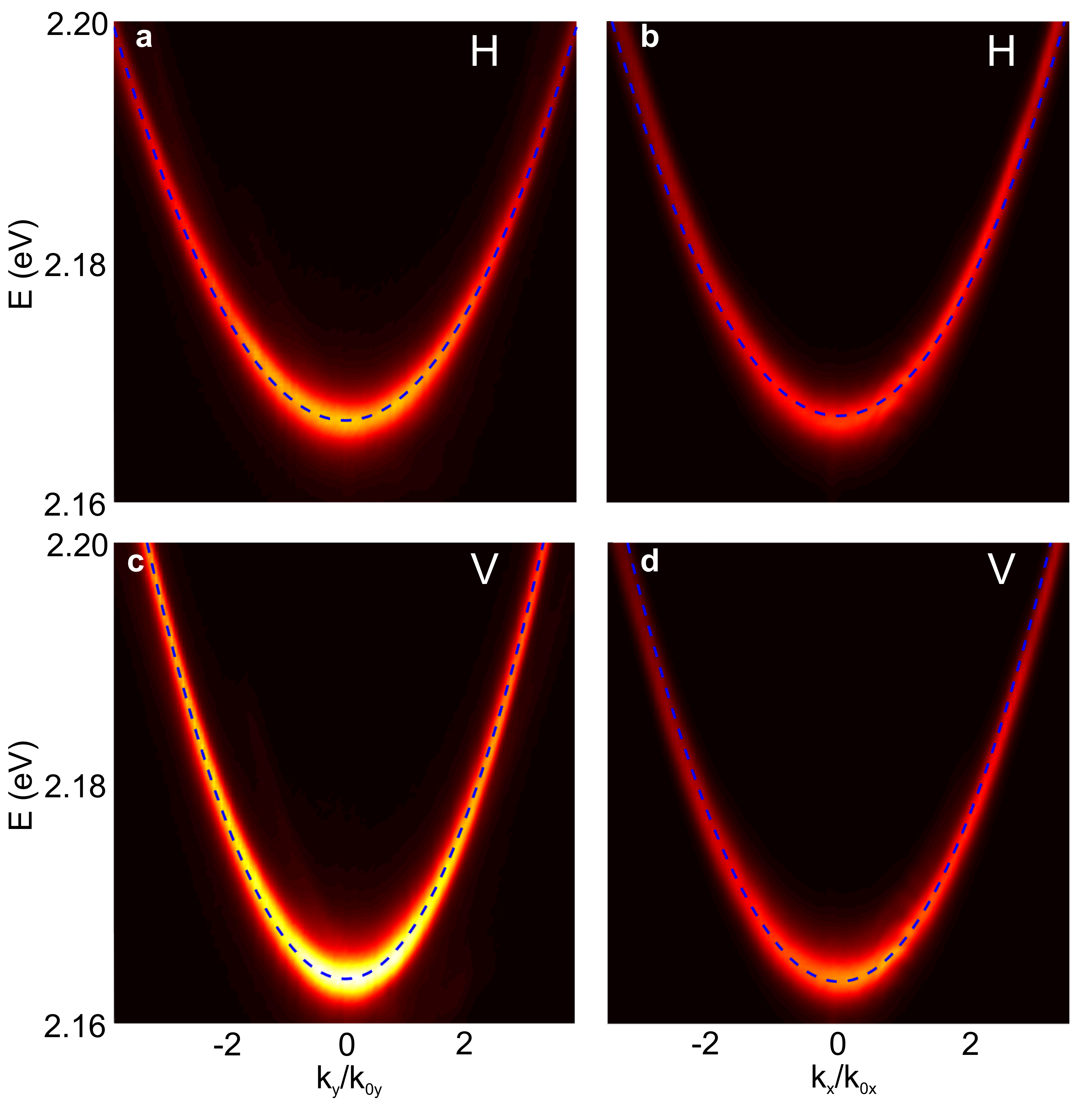}
\caption{\textbf{Fit of the dispersions.} Fit of the polarization-resolved dispersions in the two perpendicular directions. The false color map shows the reflectance as a function of wave vector and energy. Fitting allows to obtain the spin-independent masses $m_x$ and $m_y$, as well as the polarization splittings $\beta$ and $\beta'$. \label{figDisp1}}
\end{figure}
The parameters of the Hermitian Hamiltonian (1) are extracted from the fit of the dispersion in the two perpendicular directions. The fits are shown in Fig.~\ref{figDisp1}.
In a given direction, the average of the two parabolicity coefficients (inverse masses) of the two polarizations, provides the value of $1/m_x$ and $1/m_y$. Then, the difference between the two parabolicites in the two polarizations is $\beta+\beta'$ for $Y$ direction and $\beta-\beta'$ for $X$ direction. Fitting both allows to extract $\beta$ and $\beta'$ separately.

\subsection{Non-Hermitian parameter}

\begin{figure}
    \centering
    \includegraphics[width=0.8\linewidth]{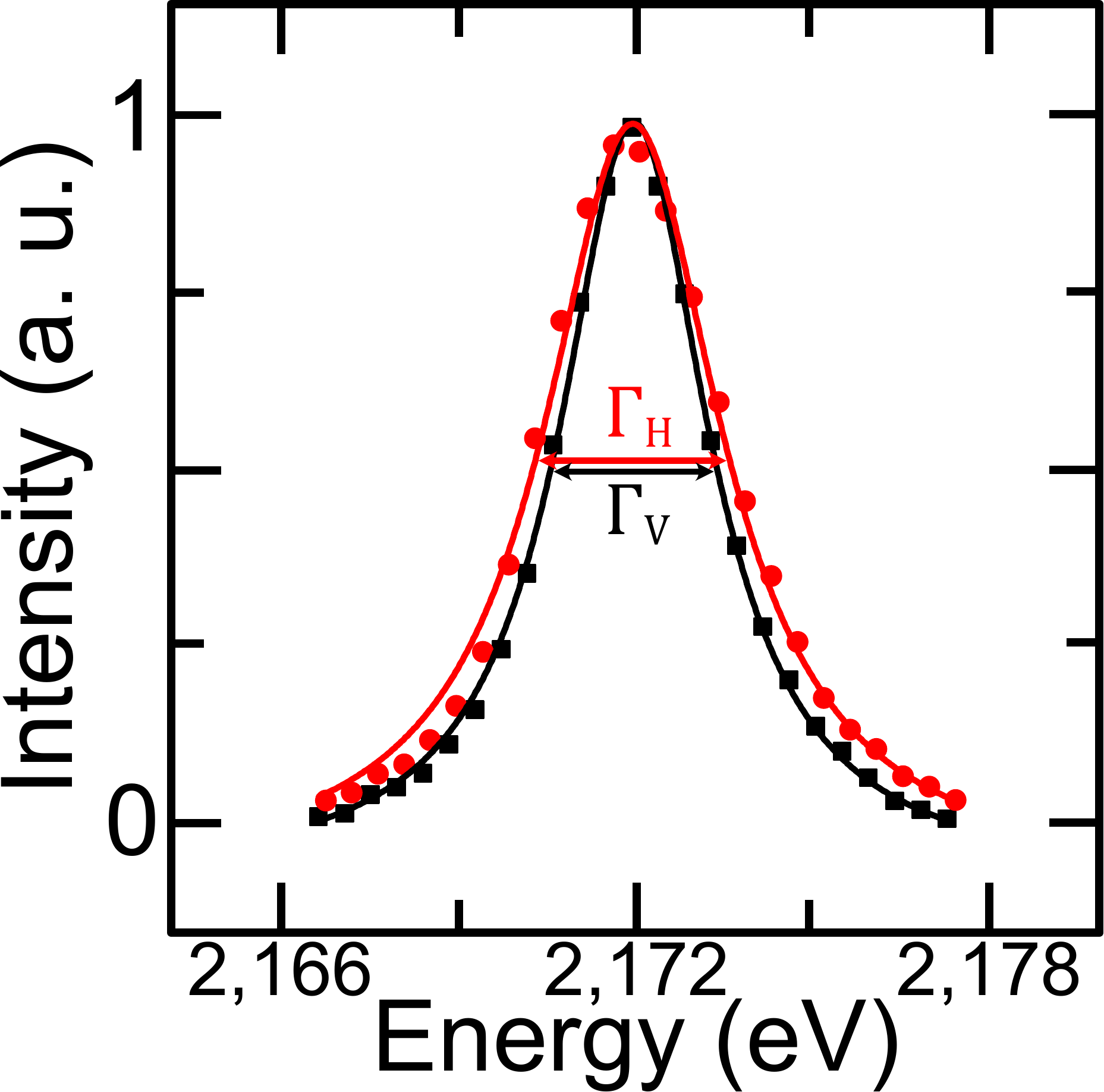}
    \caption{\textbf{Polarization-resolved reflectance spectrum.} The reflectance in polarizations H and V, measured at the position of the diabolical point and plotted as a function of energy. The difference of linewidths is approximately $13\%$.}
    \label{DGamma}
\end{figure}

In Fig.~\ref{DGamma}, we show an example of a spectrum of reflectance in two polarizations, clearly demonstrating different linewidths. The experimental points for each of the two polarizations are fitted with the Voigt function, as explained in Methods. We stress that while the non-Hermiticity of the Hamiltonian is a single constant, the linewidth at each particular wave vector is determined by the imaginary part of the corresponding eigenstate. Only at the positions of the Dirac points of the Hermitian Hamiltonian does the linewidth of the eigenstates correspond directly to the non-Hermiticity of the Hamiltonian $\delta \Gamma$. It is also the maximal possible imaginary part of the energy that could be observed for any wave vector. This is what we use to extract the non-Hermitian parameter $\delta\Gamma$.

\subsection{Hermitian winding and annihilation of exceptional points}

We consider a two-band system described by a Hamiltonian depending on a two-dimensional wave vector. The Hermitian part of the Hamiltonian is characterized by a presence of diabolical (Dirac) points. The two-band nature allows writing the Hamiltonian as a superposition of Pauli matrices. For many typical situations, such Hamiltonians can be written using only two Pauli matrices that we can choose to be $\sigma_x$ and $\sigma_y$ (massless Dirac Hamiltonian, Rashba and Dresselhaus spin-orbit couplings, TE-TM spin-orbit coupling for photons), whereas the third Pauli matrix $\sigma_z$, responsible for a symmetry breaking, opens the gap at the diabolical points (the mass term in the Dirac Hamiltonian or the Zeeman splitting for electrons and photonic modes). In this case, it is possible to characterize each of the diabolical points and the whole reciprocal space by winding numbers, which, in turn, determine the topology (the Chern number) of the bands once the gap is opened.

We then consider the evolution of the Hamiltonian with an addition of non-Hermiticity, described by a single constant parameter (independent of the wave vector). Without loss of generality, we consider the non-Hermitian contribution to be described by the Pauli matrix $\sigma_x$. In presence of the non-Hermiticity, each of the diabolical points gives rise to two exceptional points.

Thanks to the use of the Pauli matrices, it is possible to map the system's Hamiltonian to a spin-$1/2$ in a magnetic field, using the so-called pseudospin formalism, where the terms of the Hamiltonian are considered as an effective field. In the presence of a non-Hermiticity, this effective field has both real and imaginary components, $\bm{\Omega}=\bm{\Omega}'+i\bm{\Omega}''$. An exceptional point requires the real and imaginary parts of the effective field to be equal to each other in magnitude and perpendicular in direction: $\Omega'=\Omega''$ and $\bm{\Omega}'\perp \bm{\Omega}''$.

\begin{figure}[tbp]
\includegraphics[width=1.05\linewidth]{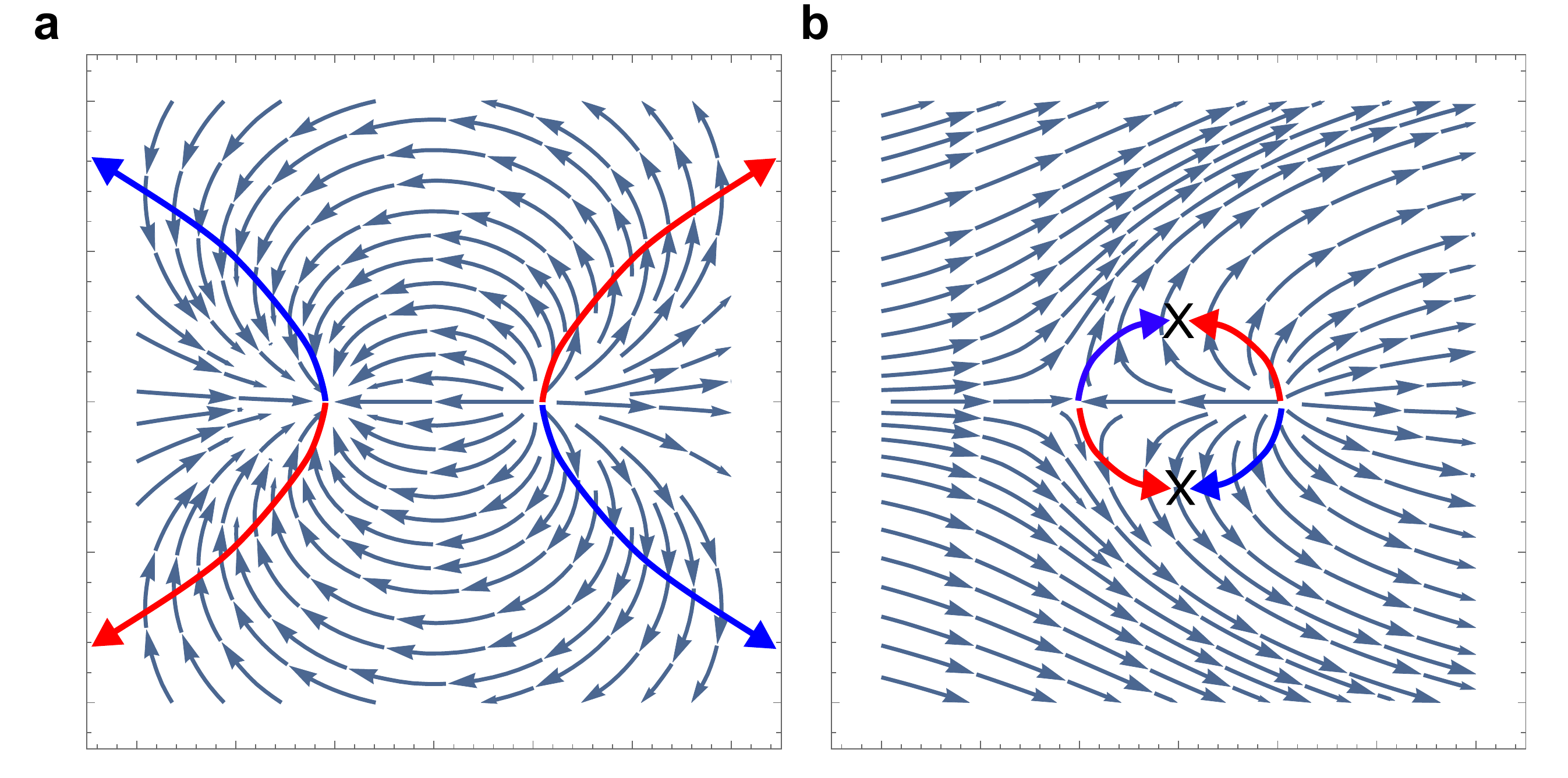}
\caption{\textbf{Hermitian winding and annihilation of exceptional points.} (a) Non-zero total winding of the real field, the exceptional points do not annihilate. (b) Zero total winding of the real field, annihilation of exceptional points. The trajectories of the exceptional points are shown with red and blue arrows (corresponding to the winding of the exceptional points themselves). Crosses mark the annihilation points. \label{Sfig1}}
\end{figure}

Since in our case the direction of the imaginary field is fixed $\bm{\Omega}''=\Omega''\bm{e}_x$, and the real field is in the XY plane, the exceptional points can only be found where the x-component of the real field is zero $\bm{\Omega'}=\Omega'\bm{e}_y\Leftrightarrow\Omega'_x=0$. This condition constrains their evolution in the reciprocal space when the imaginary field is increased $\Omega''\to\infty$ (or the real field is decreased $\Omega'\to 0$). The location of the ensemble of the points $\Omega'_x=0$ is determined by the winding of the real field at the diabolical points. With the increase of $\Omega''$, the exceptional points are moving towards infinity. \textbf{If no points, where the condition for the real part $\Omega'_x=0$ is satisfied, can be found at infinity, then the exceptional points must annihilate for some finite critical value $\Omega''_c$.} Two cases are possible:
\begin{enumerate}
    \item The overall winding of the real field is a non-zero integer. In this case, the orientation angle $\theta_r$ of the real field at infinity changes at least once between $\theta_r=0$ and $\theta_r=2\pi$, necessarily going through $\pi/2$ and $3\pi/2$, where $\Omega'_x=0$. Therefore, the exceptional points can be found for any value of $\Omega''$ up to infinity, and thus they cannot annihilate (or if they annihilate accidentally, they must then reappear when $\Omega''$ is increased).
    \item The overall winding of the real field is zero. In this case, the orientation angle of the real field $\theta_r$ at infinity is constrained to an interval which may not include $\pi/2$ or $3\pi/2$. In this case, the exceptional points cannot be present at infinity, and therefore they must annihilate for a finite value of $\Omega''_c$.
\end{enumerate}

The two situations are illustrated in Fig.~\ref{Sfig1}. Panel (a) shows the case with total winding 2 with a dipolar texture typical for the TE-TM field. The trajectories of the exceptional points are shown with red and blue lines, which are defined by the condition $\Omega'_x=0$ (vertical orientation of the real field). At a large scale, the relative position of the two diabolical points does not play any role and the texture is given by $\theta_r=2\varphi$ (where $\varphi$ is the polar angle of the wave vector). The exceptional points at infinity  are  therefore necessarily located at $\varphi=\pm\pi/4, \pm3\pi/4$ (while initially they move from the diabolical points in the vertical direction).

The second situation is illustrated by Fig.~\ref{Sfig1}(b), where the overall winding of the real field is zero. The angle of the real field at infinity is $\theta_r=0$ and there can be no exceptional points. The condition $\Omega'_x=0$ (vertical real field) gives the trajectories of the exceptional points shown in the figure, with the annihilation points marked with  crosses.

Finally, the winding of the exceptional points themselves is shown with a color of the arrows (red and blue).

The winding of an exceptional point is the winding of the phase of the complex eigenenergies. Indeed, the energy of a second-order exceptional point is given by
\begin{equation}
    E=\pm\sqrt{\alpha q}e^{iw\phi/2}
\end{equation}
where $w=\pm 1$ is precisely the winding number and $q$ is the absolute value of the parameter controlling the deviation from the exceptional point, for example, the wave vector (measured from the exceptional point). This eigenenergy can be obtained as a solution for different Hamiltonians, with different variation of real and imaginary effective fields.
Below, we show that when the imaginary effective field $\bm{\Omega}''$ is constant, the winding of the exceptional point can be easily determined from the texture of the real effective field $\bm{\Omega}'$. The most general form of a Hamiltonian of such type is
\begin{eqnarray}
    \hat{H}&=&\alpha q \left(\sigma_x\cos\varphi+iw\sigma_y\sin\varphi\right)\\
    &+&a(\left(\sigma_x\cos\theta+\sigma_y\sin\theta\right)\nonumber\\
    &+&ia\left(\sigma_x\cos(\theta\pm\pi/2)+\sigma_y\sin(\theta\pm\pi/2)\right)\nonumber
\end{eqnarray}
where the $a$ gives the values of the real and imaginary effective field oriented at an angle $\theta$ and $\theta+\pi/2$ respectively, equal and perpendicular at the exceptional point, $w$ is the winding of the parameter-dependent part of the real effective field and $\alpha$ is its strength. The eigenenergies of this Hamiltonian are given by 
\begin{equation}
    E\approx\pm\sqrt{q}\sqrt{2\alpha a}e^{\pm iw\varphi/2}e^{-i\theta/2}
\end{equation}
We see that the winding of an exceptional point is determined by the winding $w$ of the original diabolical point, modified by the relative angle of the real and imaginary fields at each exceptional point $\pm\pi/2$.


\end{document}